\documentclass[oneside, a4paper, onecolumn, 11pt]{article}

\usepackage{color}
\usepackage{graphicx}
\newcommand{\colb}{\color{black}}
\newcommand{\colr}{\color{black}}
\newcommand{\colg}{\color{black}}
\newcommand{\colm}{\color{black}}
\usepackage{caption} 

\newcommand{\2}{\!\!\!&}

\usepackage[left=2cm,top=2cm,bottom=1.5cm,right=2cm]{geometry}
\usepackage{hyperref}
\usepackage[utf8]{inputenc}
\usepackage{graphicx} 		
\usepackage{amsmath}  		
\usepackage{eurosym}
\usepackage{cite}
\usepackage{tikz}
\usepackage{chngcntr}
\usepackage{multirow}
\usetikzlibrary{trees}
\usetikzlibrary{decorations.pathmorphing}
\usetikzlibrary{decorations.markings}
\usepackage{cancel}
\usepackage{soul}

\usepackage{empheq}
\usepackage{framed}
\usepackage{feynmp-auto}
\usepackage{subcaption}
\newcommand{\bk}{\pmb{k}}

\usepackage{enumitem}
\setitemize{noitemsep,topsep=0pt,parsep=0pt,partopsep=0pt,leftmargin=*}
\usepackage{amssymb}

\usepackage{enumitem}

\usepackage{fancyhdr}

\newcommand{\vv}{\mathbf{v}}
\newcommand{\vxi}{\mathbf{p} }
\newcommand{\ren}{\mathrm{r}}
\newcommand{\reals}{\real_{\pm}}

\newcommand{\vA}{\mathbf{A}}
\newcommand{\vP}{\mathbf{P}}
\newcommand{\rel}{\mathrm{rel}}
\newcommand{\dd}{3}
\newcommand{\hel}{\mathcal{K} }

\newcommand{\he}{g}
\newcommand{\as}{\mathrm{as}}

\newcommand{\veps}{\mathbf{\eps}}

\newcommand{\mrm}{\mathrm}
\renewcommand{\mathbf}{\boldsymbol}

\newcommand{\bv}{\mathbf{v}}

\newcommand{\inc}{\mathrm{in}}

\newcommand{\bE}{\boldsymbol{E}}

\newcommand{\bn}{\mathbf{n}}

\newcommand{\bp}{\mathbf{p}}
\newcommand{\bP}{\mathbf{P}}

\newcommand{\bx}{\mathbf{x}}
\newcommand{\res}{\restriction}

\newcommand{\out}{\mathrm{out}}

\newcommand{\mcC}{\mathcal C}

\renewcommand{\i}{\mathrm i}

\newcommand{\E}{E}

\newcommand{\nr}{\mathrm{nr}} 

\newcommand{\cc}{\mathrm{c} }

\newcommand{\vx}{\mathbf{x}}

\newcommand{\ti}{\tilde}

\newcommand{\Om}{\Omega}
\newcommand{\ga}{\gamma}

\newcommand{\vk}{\mathbf{k}}

\newcommand{\be}{\beta}

\newcommand{\pa}{\partial}

\newcommand{\ov}{\overline}
\newcommand{\vp}{\mathbf{p}}   
\newcommand{\mfh}{\mathfrak{h}}

\newcommand{\eps}{\varepsilon}

\newcommand{\e}{\mathrm{e}}

\newcommand{\pho}{\mathrm{ph}}

\newcommand{\nin}{\noindent}
\newcommand{\si}{\lambda}

\newcommand{\nat}{\mathbb{N}}
\newcommand{\hil}{\mathcal{H}}

\newcommand{\mfa}{\mathfrak{A}}
\newcommand{\mco}{\mathcal{O}}

\newcommand{\fr}[2]{\frac{#1}{#2}}
\newcommand{\al}{\alpha}
\newcommand{\real}{\mathbb{R}}

\newcommand{\la}{\lambda}
\newcommand{\non}{\nonumber}
\newcommand{\Ga}{\Gamma}

\newcommand{\lan}{\langle}
\newcommand{\ran}{\rangle}

\newtheorem{theoreme}{Theorem } [section]
\newtheorem{proposition}[theoreme]{Proposition}
\newtheorem{lemma}[theoreme]{Lemma}
\newtheorem{definition}[theoreme]{Definition}
\newtheorem{corollary}[theoreme]{Corollary}
\newtheorem{remark}[theoreme]{Remark}
\newtheorem{example}[theoreme]{Example}
\newtheorem{criterion}[theoreme]{Criterion}
\newtheorem{conjecture}{Conjecture}
\newtheorem{assumption}{Assumption}

\newcommand{\bea}{\begin{assumption}}
	\newcommand{\eea}{\end{assumption}}
\newcommand{\beco}{\begin{conjecture} }
	\newcommand{\eeco}{\end{conjecture} }
\newcommand{\beq}{\begin{equation}}
	\newcommand{\eeq}{\end{equation}}
\newcommand{\beqa}{\begin{eqnarray}}
	\newcommand{\eeqa}{\end{eqnarray}}
\newcommand{\ben}{\begin{arabicenumerate}}
	\newcommand{\een}{\end{arabicenumerate}}
\newcommand{\bex}{\begin{example}}
	\newcommand{\eex}{\end{example}}
\newcommand{\ber}{\begin{remark}}
	\newcommand{\eer}{\end{remark}}
\newcommand{\bec}{\begin{corollary}}
	\newcommand{\eec}{\end{corollary}}
\newcommand{\bep}{\begin{proposition}}
	\newcommand{\eep}{\end{proposition}}
\newcommand{\becr}{\begin{criterion}}
	\newcommand{\eecr}{\end{criterion}}

\newcommand{\psin}{b}

\def\bel{\begin{lemma}}
	\def\eel{\end{lemma}}
\def\bet{\begin{theoreme}}
	\def\eet{\end{theoreme}}
\def\bed{\begin{definition}}
	\def\eed{\end{definition}}

\newenvironment{itemize*}%
  {\begin{itemize}%
    \setlength{\itemsep}{0pt}%
    \setlength{\parskip}{0pt}}%
  {\end{itemize}}

\usepackage{enumitem}

\begin{document}
\title{Infrared problem in quantum electrodynamics} 
\author{Pawe\l \, Duch\, \ and \
Wojciech Dybalski\,  \\[5mm]
\normalsize  Faculty of Mathematics and Computer Science, 
\normalsize Adam Mickiewicz University, \\
\normalsize ul.~Uniwersytetu Pozna\'nskiego 4, 61-614 Pozna\'n, Poland}
\date{}

\maketitle
\begin{abstract}
The infrared problem  in quantum electrodynamics consists of intriguing difficulties in scattering
theory appearing at large scales and low energies.  Although they can be  circumvented using  ad hoc recipes, such as 
the inclusive collision cross sections, there have been continuing efforts to reach a conceptually clear and mathematically
rigorous  understanding.
In this article we focus on such  developments of the last two decades. We start from the Buchholz-Roberts
approach in the setting of algebraic QFT, which is based on the idea that there should be no infrared problems inside the
future lightcone. Then we move on to the setting of non-relativistic QED, where insights from Haag-Ruelle scattering theory 
and the Faddeev-Kulish formalism suggest concrete formulas for the physical electron. Finally, in a 
setting of perturbative QFT, we outline a recent proposal for an infrared finite scattering matrix, which is also 
in the spirit of  the Faddeev-Kulish approach.

\end{abstract}
\begin{center} 
{\textbf{Keywords}} \\ \end{center} 
Algebraic QFT; non-relativistic QED; perturbative QED; scattering theory; M\o ller operators; scattering matrix;  LSZ reduction formulas;   Dollard formalism; Faddeev-Kulish approach;  infraparticles

\begin{center}
{\textbf{Key points}}
\begin{enumerate}

\item Infrared problem in high energy  physics.

\item Infrared problem in algebraic QFT.

\item Infrared problem in non-relativistic QED.

\item Infrared problem in perturbative QED.

\end{enumerate}
\end{center}

\section{Introduction. Infrared problem in {\colg high energy physics}}

 Given tremendous resources invested in the construction and operation of  particle colliders 
 there is a particular need to put the relevant computational procedures of quantum field theory (QFT) on a solid mathematical and conceptual basis.  
 The inherent theoretical difficulties in QFT can be divided into ultraviolet (UV) and infrared (IR) problems, according to their location on the energy scale.    
 The infrared difficulties, which we address in this article,  pertain to a  complicated scattering theory of light and electrically charged matter in quantum electrodynamics (QED). Their current understanding is only fragmentary:
On the one hand, there is a well tested computational algorithm in the setting of perturbative QED, due to Yennie, Frautschi and Suura (YFS) \cite{YFS61, We}, which is used everyday to determine  collision cross sections of physical processes.  While the YFS algorithm explains well the experimental data, it lacks a satisfactory justification from  first principles. On the other hand there is a large body of conceptually and mathematically clear work on  the infrared problem in QED in various {\colr  approaches} \cite{Ha, Str, Sp,St}. But these results are typically too abstract to compare them with experiments.  One can  speculate, that  a future  solution of the infrared problem in QED will be a bridge linking  the conceptual and computational side. We remark that a need for  better conceptual understanding of infrared problems is also recognised
in the theoretical physics community. This is demonstrated by a recent  surge of activity related to the Strominger's `infrared triangle' \cite{Stro} {\colg (see~\cite{He17} for a mathematically rigorous exposition of related ideas)}.

Let us first  briefly address  the computational aspects of the infrared problem.
We recall that QED is formally given by the Lagrangian density: 
\beqa
\mathcal{L}=\ov{\psi}(\i \ga^{\mu}(\pa_{\mu}+\i e A_{\mu})-m)\psi-\fr{1}{4}F_{\mu\nu} F^{\mu\nu}, \label{QED}
\eeqa
where $\ga^{\mu}$ are the Dirac matrices, $\psi$, $m$, ${\colr -e}$  are the Dirac field, mass and charge of the electron, $A_{\mu}$ is the electromagnetic potential and $F_{\mu\nu}:=\pa_{\mu}A_{\nu}-\pa_{\nu}A_{\mu}$  the Faraday tensor. After quantization $\psi$ and $A_{\mu}$ are  operator valued distributions on a certain Hilbert space $\hil$ {\colm and
the interaction Hamiltonian is formally given by the expression
\beqa\label{eq:QED_interaction_vertex}
V:=\int_{\real^3} d\bx\, :\!J^{\mu}(0,\vx)A_{\mu}(0,\vx)\!:, \quad J^{\mu}(x):= \ov{\psi}(x) \ga^{\mu}  \psi(x), \quad x:=(t,\vx),
\label{Interaction}
\eeqa
{\colr where $:\ldots :$ denotes normal ordering.} 
Quantum mechanics gives the following candidate for the scattering matrix ($S$-matrix) of QED
\beqa
S=\mathrm{Texp}\bigg(-\i e \int_{-\infty}^{\infty}d\tau\, V^{\mathrm{I}}(\tau)\bigg), \label{S-matrix}
\eeqa  
which we interpret as an operator on the subspace of $\hil$ spanned by the outgoing asymptotic states. Here $\mathrm{Texp}$ denotes the time ordered exponential and $V^{\mathrm{I}}$ is the interaction Hamiltonian (\ref{Interaction}) in the interaction picture. In order to make sense of the above definition of the $S$-matrix one has to overcome several problems. On the one hand, the interaction Hamiltonian $V$ exhibits ultraviolet problems due to a pointwise multiplication of distributions. {\colr This}, however,  can be resolved by renormalization, at least  order by order  in perturbation theory. On the other hand, the $S$-matrix suffers from infrared  problems due to the integration over whole space in the definition of $V$ and integration in time over entire real line in (\ref{S-matrix}). In QFT models with only massive particles the above infrared problems can be dealt with by the application of the LSZ procedure or adiabatic switching of the interaction. However, in QED due to the zero mass of the photon some persistent divergences appear, which cannot be removed by renormalization. Let us discuss them in more detail following \cite{We,YFS61}.}

\begin{figure}[t]
\centering
\begin{subfigure}{.3\textwidth}
  \centering
\begin{fmffile}{gluon0}
\begin{fmfgraph}(90,60) 
\fmfleft{i1,i2}
\fmfright{o1,o2}
\fmf{fermion}{o1,v1,v2,o2} 
\fmf{photon}{i1,v1} 
\fmf{photon}{v2,i2}
\end{fmfgraph}
\end{fmffile}
\caption{Leading contribution}
\label{leading}
\end{subfigure}
\begin{subfigure}{.3\textwidth}
\centering
\begin{fmffile}{gluon1}
\begin{fmfgraph}(90,60) 
\fmfleft{i1,i2}
\fmfright{o1,o2}
\fmf{photon}{i2,v2}
\fmf{fermion}{o1,v1,v2} 
\fmf{plain, tension=3}{v2,v3}
 \fmf{fermion, tension=3}{v3,v4}
 \fmf{plain, tension=3}{v4,o2}
\fmf{photon,left,tension=0}{v3,v4}
\fmf{photon}{i1,v1} 
\end{fmfgraph}
\end{fmffile}
\caption{Radiative correction}
\label{radiative}
\end{subfigure}
\begin{subfigure}{.3\textwidth}
  \centering
\begin{fmffile}{gluon2}
\begin{fmfgraph}(90,60) 
\fmfleft{i1,i2}
\fmfright{o1,o2}
\fmf{photon}{i2,v2}
\fmf{fermion}{o1,v1,v2} 
\fmf{plain, tension=3}{v2,v3}
 \fmf{fermion, tension=3}{v3,v4}
 \fmf{plain, tension=3}{v4,o2}
\fmf{photon}{i1,v1} 
\fmffreeze
\fmftop{o3}
\fmf{photon}{o3,v3} 
\end{fmfgraph}
\end{fmffile}
\caption{Soft photon emission}
\label{soft}
\end{subfigure}
\caption*{\small{Fig. 1. Representative Feynman diagrams entering into the computation of inclusive collision cross sections of Compton scattering. Solid lines symbolize  the electron, wiggly lines the photons. The  time axis is directed  from the bottom to the top of each figure.}  
}
\end{figure}
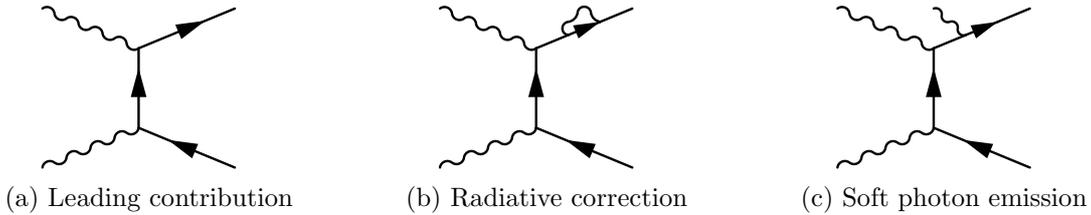

The  collision cross section $\sigma$ of a physical process evolving from an initial state~$\al$ to a final state $\be\neq \al$  is proportional to $|S_{\al,\be}|^2$, where $S_{\al,\be}$ is the corresponding $S$-matrix element. Thus schematically we can write
\beqa
\sigma\sim |S_{\al,\be}|^2. \label{S-conventional}
\eeqa
These matrix elements are computed perturbatively,  by expanding the exponential in (\ref{S-matrix}) into a power series.  The resulting formula can be expressed in terms of   Feynman diagrams, which capture the intuitive meaning of various contributions. For example, let us consider Compton scattering: both its initial state ($\al$) and final state ($\be$) contains one electron and one  photon.  The leading contribution to $S_{\al, \be}$ is given by `tree diagrams' like {\colg  in} Figure~\ref{leading}.  
Higher order corrections are given by `loop diagrams' which involve emission and reabsorption of  virtual particles.  In Figure~\ref{radiative} we illustrate one such diagram describing a \emph{radiative correction}, which is 
due to the emission and reabsorption of a \emph{soft photon} (i.e. a photon of low energy). Radiative corrections contain infrared divergencies which
have to be regularized with the help of an infrared cut-off $\si$. Such a cut-off simply eliminates photons of energy lower than $\si$ from the expression. After the regularization one obtains a fairly explicit expression for the regularized  $S$-matrix element 
$S^{\si}_{\al,\be}$ by formally
summing up the classes of diagrams containing all possible radiative corrections. However, the net effect of the divergencies of  individual diagrams
leads to vanishing  of $S^{\si}_{\al,\be}$ in the limit $\si\to 0$. Thus, by application of  standard rules of the game from quantum mechanics, we obtain a result which is unacceptable from the experimental point of view:
\beqa
\sigma\sim \lim_{\si\to 0}|S^{\si}_{\al,\be}|^2=0. \label{vanishing}
\eeqa
A pragmatic solution, proposed  by Yennie, Frautschi and Suura \cite{YFS61}, means a serious deviation from these rules of the game: 
We should not consider the process $\al\to\be$  alone, 
but rather study a whole family of processes $\al\to\be_n$, $n=0,1,2\ldots$, where $\be_n$ involves the emission of $n$ soft photons of total
energy $E$ in addition to the `hard particles' contained in  $\be$. (For example,  in the case of Compton scattering, a contribution to $\al\to\be_1$ is depicted on Figure~\ref{soft}). The resulting \emph{inclusive} collision cross section $\sigma_{\mathrm{inc}}(E)$ is schematically given by
\beqa
\sigma_{\mathrm{inc}}(E)\sim \lim_{\si\to 0} \sum_{n=0}^{\infty}  |S^{\si}_{\al, \be_n}|^2, \label{inclusive}
\eeqa 
and turns out to be finite and not identically zero. It is consistent with experimental results provided $E$ is chosen below the sensitivity of the detector so that the soft photons escape detection. Formula~(\ref{inclusive}) is a useful computational algorithm, but its rigorous derivation from a conceptually clear idea is missing.
This situation triggered a lot of  interesting work in mathematical physics, some of  which will be outlined in this article. {\colr We stress, however, that no systematic review of the literature is attempted here.}

There is a methodological difficulty present in any rigorous work on QED which should be made clear from the beginning: Due to its very singular UV structure (so called Landau pole) the full-fledged QED is not available and one has to resort to indirect approaches.  The setting of the above discussion of the YFS algorithm and of  Sec.~\ref{perturbative-section} is  perturbative QED, in which the physically relevant quantities are given by formal power series in the coupling constant,
whose convergence or summability is out of control \cite{St,DF99}.  In Sec.~\ref{AQFT-section} we work in the axiomatic setting of algebraic QFT  which provides a list of properties that a hypothetical full QED should satisfy and studies their consequences. Most conceptual results on the infrared problem are stated in this language.
A  middle course between these two approaches 
is  non-relativistic QED \cite{Sp} covered in Sec.~\ref{nrQED-section}. It deals with  mathematically well defined models obtained from the formal  QED Hamiltonian by imposing a fixed ultraviolet cut-off and removing terms responsible for the electron-positron pair creation.  These models are consistent with the conceptual knowledge about QED from the axiomatic approach and proved concrete enough to yield rigorous results comparable with experiments \cite{BFP07}. 

\section{{\colr Infrared problem in}  algebraic QFT} \label{AQFT-section}
\setcounter{equation}{0}

A relativistic algebraic QFT is defined by a net of local $C^*$-algebras of observables $\real^4\supset \mco\mapsto \mfa(\mco)\subset B(\hil)$, its $C^*$-inductive limit $\mfa$, and the energy-momentum operators $(H_{\rel}, \vP_{\rel})$ on a Hilbert space $\hil$. These objects  satisfy the standard
Haag-Kastler postulates of isotony, locality, covariance under translations and positivity of energy  {\colb for which we refer to \cite{Ha} or to a separate contribution to this Encyclopaedia \cite{E3}}. By looking at sequences of observables localized in regions shrinking to
a point one can recover pointlike localized quantum fields of the theory. In QED these
include the Faraday tensor $F^{\mu\nu}$ and the electric current {\colm$J^{\mu}$}. The electric charge is formally given by
\beqa
Q:=\int_{\real^3} d\bx\, {\colm J^0}(0,\vx),
\eeqa 
and we start the analysis of the theory with states of zero charge. One such state is the vacuum $\Om$
describing the empty space. The subspace $\hil_0:=\ov{\mfa \Om}$ is called the vacuum sector, and we will
treat it as the defining representation of QED in the following.  It contains states of electrically neutral excitations as for example photons and atoms\footnote{As we treat here atoms from the point of view of QED, we disregard the baryon number.}. Their masses are eigenstates of the relativistic mass operator $M:=\sqrt{H_{\rel}^2-\vP_{\rel}^2}$, i.e. they are particles in the sense of Wigner.

\subsection{Scattering of atoms and photons}

\nin Scattering theory for photons in the setting of algebraic QFT was developed by Buchholz in \cite{Bu77}, exploiting the Huyghens principle. Scattering theory of atoms in the presence of photons  was developed   in \cite{Dy05, He14, Du16} 
along the lines of  Haag-Ruelle theory \cite{Ha}. {\colb As it is discussed in more detail elsewhere in this Encyclopaedia  \cite{E2},  we can be brief here.}

Let $A\in \mfa$ be a local observable and $f$ be a suitable solution of the Klein-Gordon  equation.
 Then the outgoing asymptotic field of the particle has the form
\beqa
A^{\out}:=\lim_{t\to\infty}A_t, \quad A_t:=\int_{\real^3}d\bx \, A(t,\vx)f(t,\vx), \label{A-T}
\eeqa
where $A(t,\vx):=\e^{\i (H_{\rel}t-\vP_{\rel}\cdot \vx) }A  \e^{-\i (H_{\rel}t-\vP_{\rel} \cdot \vx) }$ and the limit is taken in a suitable  Ces\`aro mean. A~scattering state describing $\ell$ neutral particles (atoms or photons) has the form
\beqa
\Psi^{\out}=A_1^\out\ldots A_{\ell}^\out\Om, \label{HR-theory}
\eeqa  
where the corresponding wave packets contain only positive-energy parts, so that $A_i^{\out}$ are asymptotic creation operators.
The case of incoming scattering states $\Psi^{\inc}$ is treated analogously by taking the limit $t\to-\infty$. The  
resulting $S$-matrix element, given by 
$S_{\al,\be}=\lan \Psi^{\out}_{\al}, \Psi^{\inc}_{\be} \ran$, is related to collision
cross sections in the usual way indicated in (\ref{S-conventional}). 


\subsection{\colr Scattering of electrons and photons}\label{sec:scattering_electrons_photons}
Many infrared difficulties in QED have their origin in the fact that the spacelike asymptotic flux of the electric field 
\beqa
\phi(\bn)=\lim_{r\to\infty} r^2\bn\cdot \bE(r\bn), \quad \bn\in S^2
\label{fluxes}
\eeqa
commutes with all local observables \cite{Bu82}.  The flux is an arbitrary function on the unit sphere $S^2$, constrained only by the Gauss Law.  Thus each value of the electric charge corresponds to uncountably many disjoint irreducible representations of the algebra of observables, which may be of physical interest.  Consequently, the standard Doplicher-Haag-Roberts (DHR) 
theory of superselection sectors \cite{Ha} does not apply. Moreover, for non-zero charges  these representations cannot be Poincar\'e covariant, since the existence of $\phi$ is in conflict with unitary action of Lorentz transformations.
Similar considerations lead to a conclusion that charged particles cannot have sharp masses \cite{Bu86}. 
This  \emph{infraparticle problem} invalidates the usual Haag-Ruelle scattering theory for electrically charged particles.  A charged particle is a
composite object including  a soft photon cloud  correlated with its velocity. The cloud, whose energy distribution is singular near zero,  is needed to  keep the flux constant along the time evolution. Such \emph{infraparticles} have  in fact been constructed in  concrete models of non-relativistic QED as we will discuss in Sec.~\ref{Dollard-section} below. A scattering theory for infraparticles, available in  the setting of algebraic QFT,  is the theory of particle weights, which 
is discussed {\colb elsewhere in this Encyclopaedia \cite{E2}.}  

\begin{figure}
\centering
\begin{subfigure}{.5\textwidth}
  \centering
    \begin{tikzpicture}[scale=1]
                \begin{scope}
                \draw[fill] (-2,1)  node{\footnotesize{$\phi$}};
                \end{scope} 
        \begin{scope}
                \draw[fill] (2,1) node{\footnotesize{$\phi$}};
        \begin{scope}[->]
            \draw[gray] (-2,0) -- (2,0) node[anchor=north] {};
            \draw[gray] (0,-.8) -- (0,3) node[anchor=east] {};
        \end{scope}
        \draw (-2.5,2.5) -- (0,0);
        \draw (0,0) -- (2.5,2.5);
       \end{scope}
       \draw[fill,gray] (2.2,.15) node{\scriptsize{$x$}};
       \draw[fill,gray] (.2,3) node{\scriptsize{$t$}};
       \draw[fill] (-.5,2.3) node{\scriptsize{$V_+$}};
       \shade[left color=blue!5!white,right color=blue!30!white,opacity=0.5] (0,1)--(-1.7,2.7)--(-2.5,2.5) -- (-.5,.5);
       \draw[fill] (-.9,1.4) node{\scriptsize{$\mathcal{C}^c$}};
       \draw[dashed](-1.7,2.7)--(0.5,.5);
       \draw[dashed](1.7,2.7)--(-.5,.5);
       \draw[blue] plot[variable=\t,domain=0:1.5] ({sinh(\t)},{cosh(\t)});
       \draw[fill, blue] (1.4,2) node{\scriptsize{$\mathcal{
       C}$}};
       \draw (0,-1.3) circle (8pt);
       \draw[dotted, thick,gray] (-2.3,1.3)--(-1.7,.7);
       \draw[dotted, thick,gray] (-1.7,1.3)--(-2.3,.7);
       \draw[dotted, thick,gray] (2.3,1.3)--(1.7,.7);
       \draw[dotted, thick,gray] (1.7,1.3)--(2.3,.7);
       \begin{scope}[->]
       \draw[style={decorate, decoration={snake}, draw=red}] (.2,-1.1)--(1.5,.5)node[anchor=north] {};
              \draw[style={decorate, decoration={snake}, draw=red}] (-.2,-1.1)--(-1.5,.5)node[anchor=north] {};
              \end{scope}
      \end{tikzpicture}
  \caption*{\footnotesize{Fig.~2.~(a) A hypercone~localized representation.}}
  \label{fig:sub1}
\end{subfigure}%
\begin{subfigure}{.5\textwidth}
  \centering    
  \begin{tikzpicture}[scale=1]
          \begin{scope}
                  \draw[fill] (1.5,-.8) node{\scriptsize{$\bar{\hat{A}}_{-t}$}};
          \begin{scope}[->]
              \draw[gray] (-2,0) -- (2,0) node[anchor=north] {};
              \draw[gray] (0,-.8) -- (0,3) node[anchor=east] {};
          \end{scope}
          \draw (-2.5,2.5) -- (0,0);
          \draw (0,0) -- (2.5,2.5);
         \end{scope}
         \draw[fill,gray] (2.2,.15) node{\scriptsize{$x$}};
         \draw[fill,gray] (.2,3) node{\scriptsize{$t$}};
         \draw[fill] (-.5,2.3) node{\scriptsize{$V_+$}};
         \draw[fill] (-1.5,2.1) node{\scriptsize{$\bar{\hat{A}}_t$}};
         \draw[dashed](-1.7,2.7)--(0.5,.5);
         \draw[dashed](1.7,2.7)--(-.5,.5);
         \draw[blue] plot[variable=\t,domain=0:1.5] ({sinh(\t)},{cosh(\t)});
         \draw[fill, blue] (1.4,2) node{\scriptsize{$\mathcal{C}$}};
         \draw (0,-1.3) circle (8pt);
         \draw[dotted, thick, gray] (1.8,-1.1)--(1.2,-.5);
         \draw[dotted, thick,gray] (1.2,-1.1)--(1.8,-.5);
         \begin{scope}[->]
         \draw[style={decorate, decoration={snake}, draw=red}] (.2,-1.1)--(1.5,.5)node[anchor=north] {};
         \draw[style={decorate, decoration={snake}, draw=red}] (-.2,-1.1)--(-1.5,.5)node[anchor=north] {};
         \draw[green] (-1.65,2.15)--(-2,2.5);
         \draw[red](1.6,-1.1)--(2,-1.5)node[anchor=south] {};
         \end{scope}
         \draw[green] (-1.35,1.85)--(-.8,1.3);
         \draw[red] (-.5,1)--(1.2,-.7);
        \end{tikzpicture}
  \caption*{\footnotesize{Fig.~2.~(b) (Non-)existence of asymptotic fields.}}
  \label{fig:sub2}
\end{subfigure}
\label{mass-shell-fig}
\end{figure}

\normalsize{}

There is, however, a {\colr competing} approach which we intend to outline here following \cite{BR14, AD15}:  The above discussion relies on a tacit assumption that the flux (\ref{fluxes}) exists in the considered  representations of the algebra of observables. But apart from such \emph{infraparticle representations} one can also consider  \emph{infravacuum representations} in which  the fluctuations of
the electric field tend to infinity under large spacelike translations. We remark that a distinction into these two classes of sectors can be found in a still highly recommendable review of Kraus \cite{Kr83}. More recently,  Buchholz and Roberts considered  a class of infravacuum representations motivated by the idea that infrared problems should not occur inside a future lightcone  \cite{BR14}.
Heuristically, one possibility to obtain an infravacuum representation  is to switch on a highly fluctuating background radiation, emitted in very distant past. This radiation, which is distinct from the soft photon clouds mentioned above,  should prevent the existence of the limit (\ref{fluxes}) that is `blur the flux'.  On the other hand, it is {\colr plausible} from Fig.~2\footnote{{\colg Fig. 2 is taken from \cite{AD15 } with publisher's consent.}}.~(a) and the Huygens principle that this  radiation 
{\colr should} stay outside any future lightcone~{\colr $V_+$}. 
Hence, inside $V_+$ one can follow the DHR philosophy to pass from the defining vacuum representation $\iota$ of the algebra of observables $\mfa$ to an electrically charged positive energy representation $\pi$. For this purpose, one considers a  pair of opposite charges  in a hypercone $\mcC\subset V_+$, which is a region indicated in  Fig.~2~(a). 
Next,  one shifts one of the charges to  lightlike infinity.  It is argued in \cite{BR14}, that this process of charge creation in $\mcC$ should not significantly affect operations performed in the spacelike complement of $\mcC$ in $V_+$, denoted $\mcC^\cc$. Therefore, the resulting charged representation $\pi$ should satisfy the following property of \emph{hypercone localization}
\beqa
\pi\res \mfa(\mcC^\cc)\simeq\iota \res \mfa(\mcC^\cc). 
\label{intro-hypercone}
\eeqa
Here $\simeq$ denotes unitary equivalence and $\mfa(\mcC^\cc)$ is the algebra of all observables which can be measured in $\mcC^\cc$. 
Since $\mcC^\cc$ is a subset of $V_+$, this property is not in conflict with high fluctuations of the electric field at spacelike infinity, blurring the flux (\ref{fluxes}) {\colr as shown in} Fig.~2~(a). 
As $\phi$ does not exist,
we may assume that $\pi$ is
covariant under Poincar\'e transformations and that charged particles 
have sharp masses.  For some evidence
 in favour of this latter assumption we refer to \cite{CD20}.

Using these assumptions and the existence of asymptotic photon fields
in the vacuum representation  \cite{Bu77},  the outgoing asymptotic photon fields $A^{\out}$ were constructed by Alazzawi and Dybalski  in \cite{AD15}    in  hypercone  localized 
representations from \cite{BR14}.  Moreover, the outgoing Compton scattering states were obtained. They have the form:
\beqa
\hat\Phi^{\out}=A_1^{\out}\ldots A_{\ell}^{\out}\hat\Phi^{\out}_{\mathrm{el}}, \label{Compton-axiomatic}
\eeqa
where $\hat\Phi^{\out}_{\mathrm{el}}$ is a single electron state, and its superscript `out' indicates that it was constructed in a representation localized in a \emph{future} lightcone. 
It was crucial for the construction that the approximating sequences $A_t$ of $A^{\out}$ from  formula~(\ref{A-T}) can be localized in subsets of the future lightcone, as indicated 
in Fig.~2.~{\colr  (b)}, and thus they do not interfere  with the highly fluctuating background radiation. As also shown in  the same figure, the incoming photon fields are not expected to  exist in this representation as their
approximating sequences collide with the background radiation. Thus to construct incoming Compton scattering states it is necessary
to pass to a Buchholz-Roberts representation localized in a \emph{backward} lightcone. As both representations act naturally on the
same Hilbert space, it is possible to define $S$-matrix elements.

\section{{\colr Infrared problem in} non-relativistic QED} \label{nrQED-section}
\setcounter{equation}{0}

As mentioned in the Introduction, the Hamiltonian of full QED, coming from the Lagrangian density~(\ref{QED}), is a very singular object. However, by fixing the Coulomb gauge,  introducing an ultraviolet cut-off in the interaction, 
removing terms responsible for the electron-positron pair creation, and several other simplifications dictated by convenience, one obtains various Hamiltonians of non-relativistic QED. 
These Hamiltonian capture the low-energy properties of QED  and  thus are well suited for the analysis of  infrared problems. 

A Hamiltonian of non-relativistic QED, which we consider in this article, describes massive,  spinless particles, called  electrons
or atoms (depending if they are electrically charged or not) which are coupled to the second-quantized electromagnetic field.
We denote the single electron (or single atom) Hilbert space 
 by $\hel=L^2(\real^{\dd})$ and the single photon Hilbert space by $\mfh=L^2(\reals^{\dd})$, where $\reals^3:=\real^3\times \mathbb{Z}_2$ is the
photon configuration space including the polarization degrees of freedom. The corresponding Fock spaces, denoted 
$\Ga(\hel)$, resp. $\Ga(\mfh)$, carry the creation and annihilation operators $\psin^*, \psin$, resp. $a^*, a$.  The  Hilbert space of the model has the form $\hil_{\nr}=\Ga(\hel)\otimes \Ga(\mfh)$ and the Hamiltonian is given by  \cite{Fr74.1}
\beqa
H_{\nr}:=\fr{1}{2m}\int_{\real^3} d\vx\, {\colr \psin^*(\vx)}
\big(-\i\nabla_{\vx}-{\colr e}\vA(\vx)\big)^2\psin(\vx)
+\int_{\reals^3} d\vk\, |\vk|a^*(\vk)a(\vk). \label{standard-nrQED}
\eeqa
{\colr The} first term on the r.h.s. above describes the free evolution of the massive particles and their interaction with photons, and the second
term, which we  call $H_{\pho}$, {\colr governs} the free evolution of photons.  
{\colr In particular,} $\vA$ is the second quantized electromagnetic potential in the Coulomb gauge:
\beqa
\vA(\vx)=\int_{\reals^3}d\vk \fr{\ti\rho(\vk)}{\sqrt{2|\vk|}}\{  \e^{-\i\vk\cdot \vx }\veps(\vk)a^*(\vk) + \e^{\i\vk\cdot \vx }\veps(\vk)^*a(\vk)\}. \label{A}
\eeqa
Here $\veps$ are the photon polarization vectors and $\ti\rho\in C_0^{\infty}(\real^3)$ is the Fourier transform of the charge density of the
massive particle. Thus $q:={\colr e}(2\pi)^{3/2}\ti\rho(0)={\colr e}\int d\vx\, \rho(\vx)$ is the total charge which decides if the massive particles are electrons ($q\neq 0$)
or atoms ($q=0$).   Apart from the Hamiltonian~(\ref{standard-nrQED}), we will also occasionally refer to  Nelson-type  models in which
$\vA$ is replaced with the scalar field and the interaction part in the Hamiltonian is linear in this field.

Since $H_{\nr}$ leaves the number of massive particles invariant, it can be expressed as a direct sum of Hamiltonians $H^{(n)}_{\nr}$
acting on $n$-electron (or $n$-atom) subspaces  $\hil_{\nr}^{(n)}=(\bigotimes_{\mathrm s/\mathrm a}^n\hel)\otimes \Ga(\mfh)$, where 
$\mathrm s/\mathrm a$ denotes symmetrization or anti-symmetrization depending whether the massive particle is a boson or  fermion. As the case $n=1$ attracted most attention, we set  $H:=H^{(1)}_{\nr}$,  $\hil:=\hil_{\nr}^{(1)}$ and  recall the
familiar formula \cite{Sp}
\beqa
H=\fr{1}{2m} (-\i\nabla_{\vx}{\colr-e}\vA(\vx)\big)^2+\int_{\reals^3} d\vk\, |\vk|a^*(\vk)a(\vk).
\label{PF-model} 
\eeqa

Scattering theory in  non-relativistic QED  has been thoroughly studied over the last 
two decades. The first step is a construction of asymptotic creation and annihilation operators of photons with some wave functions
$h\in \mfh$  
\beqa
a_{\out}^{(*)}(h)=\lim_{t\to\infty}\e^{\i tH_{\nr}}a^{(*)}( \e^{-\i t|\vk| }h) \e^{-\i tH_{\nr}}, \label{asymptotic-creation-operator}
\eeqa
as strong limits on  a certain domain in $\hil_{\nr}$. They correspond to expression~(\ref{A-T}) from the axiomatic setting. 
The existence of these limits is known under very
general conditions in the model (\ref{standard-nrQED}) 
and in similar Nelson-type models \cite{DG04, FGS01}. 
The second step is the construction of states from $\hil$  describing one physical massive particle: By translational invariance, the Hamiltonian (\ref{PF-model})  commutes with the total momentum operators $\vP:=-\i \nabla_{\vx} + \vP_{\pho}$, 
where  $\vP_{\pho}:=\int_{\reals^\dd}d\vk\, \vk \, a^*(\vk)a(\vk)$ are the photon momentum operators.   
Consequently, the Hamiltonian $H$ has a decomposition into fiber Hamiltonians $\{H(\vxi)\}_{\vxi\in \real^{\dd}}$ at fixed momentum $\vxi$,
which are concrete operators on $\Ga(\mfh)$, i.e.,
\beqa
H=I^*\int_{\real^{\dd}}^{\oplus} d\vxi\,H(\vxi) I, \label{fiber-decomposition}
\eeqa
where $I:\hil\to L^2(\real^3;\Ga(\mfh))$ is a suitable unitary map. 
The infimum of the spectrum of $H(\vxi)$ is denoted $E(\vxi)$ and the function $\vxi\mapsto \E(\vxi)$
is the renormalized dispersion relation of the physical massive particle. The
question  {\colg of whether} $E(\vxi)$ is an eigenvalue of $H(\vxi)$, essential for the construction of scattering states, depends  {\colg on whether} the massive particle is an atom or an electron.
These two cases will be discussed separately below. For the sake of clarity, we omit most technical assumptions in this overview.

\subsection{Scattering of atoms and photons}
\newcommand{\phin}{b}

\nin We recall that massive particles
are called atoms if $\tilde{\rho}(0)=0$, where $\tilde{\rho}$ appeared  {\colg in \eqref{A}}. In this case $E(\vxi)$ is 
a simple eigenvalue of $H(\vxi)$, corresponding to  an eigenvector $\Psi_{\vxi}$ \cite{FGS04, FGS07, AGG05, LMS07}. 
These eigenvectors can be superposed into wave packets
\beqa
\Psi_{\he}=I^*\int^{\oplus}_{\real^{\dd}}d\vxi\, \he(\vxi) \Psi_{\vxi},\quad \he\in L^2(\real^{\dd}), 
\label{ren-single-electron-states}
\eeqa   
which give physical single atom states.
Scattering states, describing one 
physical electron and $\ell$ photons, are defined by 
\beqa
\Psi^{\out}=a_{\out}^{*}(h_1)\ldots a_{\out}^{*}(h_{\ell})\Psi_{\he}. \label{scattering-states}
\eeqa
Clearly, there is certain asymmetry between  photons
and atoms in the above formula: Only for photons we use asymptotic creation operators. Construction of  asymptotic creation operators for atoms can be avoided as long as there is only one atom in the scattering process. If two or more atoms are present, 
their asymptotic creation operators are needed {\colr to define the corresponding scattering states}.  Such operators were introduced in the context of
massive theories in  \cite{Al73}  and are given by a rather complicated formula: First, we 
define the \emph{renormalized} creation operator of an atom
\beqa
\psin_{\ren}^*(\he ):=\sum_{j=0}^{\infty}\fr{1}{\sqrt{j!}}\int_{\real^3} d\vxi\,\int_{\reals^{3j}} d\vk\,\he(\vxi) f^{j}_{\vxi}(\vk_1,\ldots,\vk_j)a^*(\vk_1)\ldots a^*(\vk_j) 
\psin^*(\vxi- \sum_{i=1}^j \vk_i), \label{electron-creation-op-ren}
\eeqa
where $\{f^{j}_{\vxi}\}_{j\in\nat_0}$ are the $j$-particle components of $\Psi_{\vxi}\in \Ga(\mfh)$ and $\psin^*$  is the usual electron creation operator in momentum space. By definition, $\psin_{\ren}^*$ 
satisfies $\Psi_{\he}=\psin_{\ren}^*(\he )\Om$, where $\Om$ is the vacuum vector of $\hil_{\nr}:=\Ga(\hel)\otimes \Ga(\mfh)$
and $\Psi_{\he}$ is given by (\ref{ren-single-electron-states}). Consequently, in the limiting case of $g$ equal to the Dirac delta at $\bp$, we have
\beqa
(H - E(\bp))\psin_{\ren}^*(\bp)\Om=0. \label{ren-energy}
\eeqa
Now the asymptotic creation operator of the electron
is defined by analogy with (\ref{asymptotic-creation-operator})
\beqa
\psin_{\ren,\out}^*(\he):=\lim_{t\to\infty}\e^{\i t H_{\nr}} \psin_{\ren}^*(\e^{-\i tE(\vp)}\he )   \e^{-\i t H_{\nr}}, \label{asymptotic-electron-op} 
\eeqa
if the limit exists. Two-atom scattering states of the schematic form
\beqa
\Psi^{\out}_{2}:= \psin_{\ren,\out}^*(\he_1) \psin_{\ren,\out}^*(\he_2)\Om \label{two-electron}
\eeqa
are under control in the massless Nelson model \cite{DP14}.  A generalization to an arbitrary number of atoms and photons or to more realistic models seems to be within reach of  the existing methods, but has not been accomplished so far.

\newcommand{\D}{\mathrm{D}}
\newcommand{\II}{\mathrm{I}}

\subsection{Scattering  of electrons and photons}\label{Dollard-section}

If the massive particle is an electron, that is $\tilde\rho(0)\neq 0$, then $E(\vxi)$ is not an eigenvalue of $H(\vxi)$
except for $\vxi=0$ \cite{HH08}. Thus there are no normalizable states describing the bare electron in empty space. 
In other words,  the electron is an \emph{infraparticle}  and the scattering theory used in the case of electrons and atoms above does not apply.

{\colr
To find out how to proceed, let us recall  briefly  the  Dollard approach to quantum mechanical long-range scattering, following  \cite{Do64, Dy17}. 
Let us consider  a  particle moving in an external potential. Its Hamiltonian, acting on a Hilbert space $\ti{\hil}$, is given by 
$\ti H=\ti{H}_0+{\colr e}\ti{V}(\bx)$, where $\ti{H_0}$ is the free Hamiltonian, the potential $\ti{V}$ depends on the position 
of the particle {\colr and we use tilde to distinguish the quantum mechanical quantities from their QFT counterparts}. It is well known that for long-range  potentials the conventional
wave operators $\ti{\Om}_{\mathrm{conv}}^{\mrm{out}  }= \lim_{t\to  \infty} \e^{\i t\ti H} \e^{-\i t\ti H_0}$ 
do not exist as limits in the strong topology. The idea of  Dollard was to 
replace the free  dynamics  $\e^{-\i t\ti H_0}$ in this formula with a modified dynamics $\e^{-\i t\ti H_0}\ti{U}^{\D}(t)$,
where the Dollard modifier $\ti{U}^{\D}(t)$ is obtained as  follows:
\begin{enumerate}

\item[$\bullet$] Define the time-dependent asymptotic potential $\ti{V}_{\colr\as}(t):=\ti{V}\big(\fr{\bp}{m}t \big)$ by evaluating the potential $V(\bx)$ at the
expected ballistic trajectory of the particle $ \fr{\bp}{m}t$.

\item[$\bullet$] Transform the asymptotic potential to the interaction picture $ \ti{V}^{\mathrm{I}}_{\colr{\as}}(t):=\e^{\i  t\ti{H}_0}\ti{V} \big( \fr{\bp}{m} t \big) \e^{-\i t\ti{H}_0}$.

\item[$\bullet$] Define the Dollard modifier as the time ordered exponential $\ti{U}^{\D}(t):=\mathrm{T}\exp\big(-\i {\colr e}\int_0^td\tau\, \ti{V}^{\II}_{\colr\as}(\tau)\big)$.

\end{enumerate}
The Dollard wave operators have the form $\ti{\Om}^{\mathrm{out}  }= \lim_{t\to  \infty} \e^{\i t\ti H} \e^{-\i t\ti H_0} \ti{U}^{\D}(t)$.
They exist as  strong limits of their approximating sequences, also for physically relevant long-range potentials. To understand the above
prescription one should note that the \emph{asymptotic dynamics} $\ti{U}_{\mrm{as}}(t):=\e^{-\i t\ti H_0}\ti{U}^{\D}(t)$ satisfies
\beqa
\i\pa_{t} \ti{U}_{\mrm{as}}(t)=(\ti{H}_0+{\colr e}\ti{V}_{\colr\as}(t)) \ti{U}_{\mrm{as}}(t).
\eeqa
That is,  its Schr\"odinger  evolution is governed by the asymptotic Hamiltonian $\ti H_{\mrm{as}}:=\ti{H}_0+{\colr e}\ti{V}_{\colr\as}(t)$. }

Thus, the  scattering states are obtained as limits as $t\to\infty$ of 
\beqa
\ti\Psi_t= \e^{\i t \ti{H}}\ti{U}_{\mrm{as}}(t)\Psi_0=\int d\bp\,  h(\bp) \e^{\i t \ti{H}} \e^{-\i t\ti H_0}\ti{U}^{\D}(t)   |\,\bp\,\ran,    \label{quantum-mechanics} 
\eeqa
where $\ti\Psi_0:=\int d\bp\,  h(\bp)|\,\bp\,\ran\in \ti\hil$.  The  second step above was taken in preparation for the case below, in which the 
asymptotic dynamics is $\bp$-dependent.

Let us now try to apply this procedure to non-relativistic QED in the spirit of Faddeev and Kulish~\cite{FK70}, adapting the discussion from \cite{Dy17}. The first step of the Dollard formalism is to identify the asymptotic Hamiltonian, which captures the 
large-time dynamics of the system. For an electron, moving with group velocity $\vv_{\bp}:= \nabla E(\bp)$, a natural candidate is 
\beqa
H_{\as,\bp}(t):= H_{0,\bp}+{\colr e}V_{\as,\bp}(t), \quad {\colr V_{\as,\bp}(t):=-\vv_{\bp} \cdot \vA(\vv_{\bp} t).} \label{asympt-hamiltonian}
\eeqa 
Here $H_{0,\bp}$ is the renormalized free Hamiltonian given by
$H_{0,\bp}:=E(\bp)+H_{\pho}$, where $E(\,\cdot \,)$ is the physical dispersion relation of the electron defined below formula (\ref{fiber-decomposition})
and ${\colr V_{\as, \bp}}$ is an interaction term from (\ref{PF-model}) evaluated at the asymptotic ballistic trajectory of the particle. Aiming for simplicity of the asymptotic Hamiltonian,  we included only  the part of the interaction linear in $\vA$.
Now the Dollard modifier has the form
\beqa
U^{\D}_{\bp}(t):=\mrm{Texp}\bigg(-\i{\colr e}\int_0^{t}  \e^{\i  H_{0, \bp}  \tau} {\colr V_{\as, \bp}(\tau) }   \mathrm{e}^{{-}\i H_{0, \bp}  \tau}  d\tau\bigg) 
\eeqa
and the resulting asymptotic dynamics $U_{\as, \bp}(t)= \e^{-\i H_{0,\bp}t } U^{\D}_{\bp}(t)$ satisfies the time-dependent Schr\"odinger equation with the Hamiltonian~(\ref{asympt-hamiltonian}).  

Now the quantum mechanical formula~(\ref{quantum-mechanics}) suggests that the approximating sequence of a scattering state describing the physical electron should have the form
\beqa
\Psi_t=\int d\bp \, h(\bp) \e^{\i tH} \e^{-\i t H_{0,\vp}  } U^{\D}_{\bp}(t) |\,\bp; t \ran. \label{intermediate-psi-t}
\eeqa
Here $ |\,\bp; t \ran $ is {\colr a} certain, possibly time-dependent, family of vectors, replacing the plane wave configurations $|\,\bp\,\ran$ familiar from quantum mechanics. 
To determine $ |\,\bp; t \ran  $ we rewrite $\Psi_t$ in the spirit of Haag-Ruelle theory to impose on it the structure of a cloud of soft photons accompanying a  bare electron.
For this purpose,   we compute the Dollard modifier
\beqa
U^{\D}_{\bp}(t)= \e^{-\i C_{\bp}t } \e^{\i \ga'_t(\bp)}  W((1-\e^{\i |\vk|t-\i \vk\cdot \bv_{\bp}t })f_{\bp}), \quad   f_{\bp}(\vk):= {\colr -e}\fr{\ti\rho(\vk)}{\sqrt{2} |\bk|^{3/2}}    \fr{  P_{\mrm{tr}}(\hat{\bk})  \bv_{\bp}}{1-\hat{\bk}\cdot \bv_{\bp}}, \label{U-D}
\eeqa
where $\hat{\bk}:=\bk/|\bk|$,  $P_{\mrm{tr}}(\hat{\bk})$ is the projection on the orthogonal complement of $\bk$, we denoted by $C_{\bp}t, \ga'_t(\bp)$ certain phases   and  $W(g):=\e^{a^*(g)-a(g)}$, $g\in L^2(\real_{\pm}^3)$, are  the Weyl operators. We note that  $f_{\bp}$ is not square integrable in $\bk$, but  the Weyl operator is well defined due to the  pre-factor $(1-\e^{\i |\vk|t-\i \vk\cdot \bv_{\bp}t })=O(|\bk|t)$. In the next step we intend to use the Weyl relations to write
\beqa
W((1-\e^{\i |\vk|t-\i \vk\cdot \bv_{\bp}t })f_{\bp})=  \e^{\i \ga''_t(\bp)} W(f_{\bp}) W(-\e^{\i |\vk|t-\i \vk\cdot \bv_{\bp}t } f_{\bp} ),  \label{Weyl-relations}
\eeqa
where both Weyl operators {\colr on the r.h.s.} are ill defined as they stand. To make sense of such expressions we introduce temporarily an infrared cut-off $\la>0$ in the Hamiltonian (\ref{PF-model}).
This means, in particular, that the function $ \ti\rho$ in (\ref{U-D}) vanishes in a neighbourhood of zero. We refrain from writing this infrared cut-off explicitly in (\ref{Weyl-relations})
in order not to overburden notation. Now the integrand in~(\ref{intermediate-psi-t}) can be rewritten as follows
\beqa
\2 \2\e^{\i tH} \e^{-\i t H_{0,\vp}  } U^{\D}_{\bp}(t)   |\,\bp; t \ran  \non\\
\2 \2\phantom{3333}= \e^{\i\ga_t(\bp)}  \{\e^{\i tH} W( \e^{-\i |\vk|t}f_{\bp})  \e^{-\i tH} \}  \e^{\i (H - E(\bp))t }\big[ W(- \e^{-\i \bk\cdot \bv_{\bp}t }  f_{\bp} )  \e^{-\i H_{\pho}t } 
\e^{-\i C_{\bp}t }  |\,\bp; t \ran   \big].
\eeqa 
By comparison with (\ref{asymptotic-creation-operator}), we observe that the expression in curly bracket approximates as $t\to \infty$ the asymptotic photon field, thus can be
interpreted as {\colr a} soft photon cloud propagating along the future lightcone.  Now relation (\ref{ren-energy}) suggests, that $|\,\bp; t  \ran$ should be determined by the 
following requirement
\beqa
\big[ W(- \e^{-\i \bk\cdot \bv_{\bp}t }  f_{\bp} )  \e^{-\i H_{\pho}t } \e^{-\i C_{\bp}t } |\,\bp; t \ran\big]= \psin_{\ren}^*(\bp)\Om. \label{solving-for-Psi}
\eeqa
In fact, $\psin_{\ren}^*(\bp)\Om$ is a bare electron state, which exists  for $\la>0$ and satisfies $ \e^{\i (H - E(\bp) )t }\psin_{\ren}^*(\bp)\Om=\psin_{\ren}^*(\bp)\Om$. Altogether, we arrive at 
\beqa
\Psi_t=\int d\bp \, h(\bp) \e^{\i\ga_t(\bp)}  \e^{\i tH} W( \e^{-\i |\vk|t}f_{\bp})  \e^{-\i tH}   \psin_{\ren}^*(\bp)\Om. \label{CFP-representation}
\eeqa
Thus guided by the Dollard formalism and Haag-Ruelle theory we are led to a definite formula for the physical electron.  

The problem of convergence of $\Psi_t$ as $t\to \infty$ was solved in the Nelson model and in the model (\ref{PF-model}) of non-relativistic QED by Chen, Fr\"ohlich and Pizzo \cite{Fr73, Fr74.1, Pi05, CFP07}. These authors used a representation similar to (\ref{CFP-representation})  with a time-dependent infrared cut-off $\la_t$ s.t. $\lim_{t\to\infty} \la_t=0$. Once {\colr the} single electron problem is solved, one can add hard photons analogously to (\ref{scattering-states}). Also scattering of one electron and one atom is under control \cite{DP18}. However, construction of scattering states of several electrons appears to be very difficult in this approach. This merits {\colr a} search for more convenient formulas for the approximating vector of the physical electron.

As the presence of the time-dependent infrared cut-off in (\ref{CFP-representation}) makes the structure  of the proof of convergence  rather intricate,  one can ask if
this cut-off can be avoided. For this purpose, let us  come back to  (\ref{solving-for-Psi}) and try to solve it for $|\,\bp; t\ran$. 
If we undo in (\ref{solving-for-Psi}) the Dollard prescription by replacing  $\bv_{\bp}t$ with the electron position $\bx$, then the solution reads 
\beqa
|\,\bp; t\ran =\e^{\i H_{\pho}t } \e^{\i C_{\bp}t } \psin_{\mrm{r},\mrm{mod}}^*(\bp)\Om.
\eeqa
Here the modified renormalized creation operator involves the wave functions of the ground state $\Phi_{\bp}$ of the modified Hamiltonian 
$H^{\mrm{w}}(\bp):=W(f_{\bp})H(\bp) W(f_{\bp})^*$  obtained by a singular Bogolubov transformation $W(f_{\bp})\,\cdot \,W(f_{\bp})^*$. In contrast
to the ground state of $H(\bp)$, the ground state $\Phi_{\bp}$ of $H^{\mrm{w}}(\bp)$ exists also in the absence of the infrared regularization.
By substituting this $|\,\bp; t \ran$ back to  (\ref{intermediate-psi-t}) and, for consistency,  undoing also there the Dollard prescription, we arrive at the formula
\beqa
\Psi_t(\bx)=\e^{\i Ht} \e^{-\i \bP_{\pho}\cdot \bx} {\fr{1}{(2\pi)^{3/2}}}\int d\bp \,  h(\bp) \e^{ \i (\bp\cdot \bx-E(\bp) t) }  \e^{ \i \gamma( \bp,\bx,t)} 
 W\big( f_{\bp} (\e^{-\i |\bk|  t+\i \bk\cdot \bx}-1) \big)  \Phi_{\bp}, \label{BDG}
 \eeqa
which does not require any infrared regularization. Here  $\bx\mapsto \Psi_t(\bx)$  should be understood as a Fock space valued function in the
sense of the identification $L^2(\real_{\pm}^3)\otimes \Ga(\mfh)\simeq L^2(\real_{\pm}^3; \Ga(\mfh))$. In the Nelson model   a rather
transparent proof of convergence of (\ref{BDG}) as $t\to \infty$ was found recently by Beaud, Dybalski and Graf \cite{BDG21}. This offers a fresh look
at the problem of scattering of several electrons which, however, remains open to date.

\section{{\colr Infrared problem in} perturbative QED}\label{perturbative-section}
\setcounter{equation}{0}

\newcommand{\rmod}{\mathrm{mod}}
\newcommand{\rout}{\mathrm{out}}
\newcommand{\rin}{\mathrm{in}}
\newcommand{\ret}{\mathrm{ret}}
\newcommand{\adv}{\mathrm{adv}}
\newcommand{\ras}{\mathrm{as}}

We now turn to the discussion of the infrared problem in the setting of perturbative QED. Let us stress that, despite its perturbative nature, perturbative QED is a mathematically well-defined model of relativistic QFT that is amenable to rigorous analysis. In the setting of perturbative QED the correlation functions, both the Wightman as well as the Green functions, are formal power series in the coupling constant whose coefficients are Schwartz distributions. In the construction of the coefficients of these series one encounters the usual UV and certain mild IR divergencies. A~rigorous construction was developed in the 70's and is well-understood mathematically. Let us only mention that the UV problem is solved using the standard renormalization procedure. To address the mild IR problem due to the presence of massless photons one introduces some IR cutoff and subsequently shows that it can be removed provided certain renormalization conditions are satisfied. The solution of the above-mentioned IR problem is standard and is not the subject of our discussion. Instead, in what follows, we concentrate on the IR problem in the description of the scattering of electrically charged particles, which is related to   slow decay of the correlation functions. In contrast to the former IR problem, the later is still not fully understood. 

\subsection{Origins of IR divergencies in perturbative QED}

In massive theories the conservation of energy ensures that the number of particles emitted during scattering is always finite. In contrast, in QED nothing prohibits a production of infinitely many photons with sufficiently small energies. In fact, heuristic reasoning suggests that the probability of an emission of a low-energetic photon in a non-trivial scattering of charged particles should be proportional to the inverse of the energy of the photon. As a result, the expected number of emitted photons is generically infinite and the amplitude of the transition to a state with a finite number of particles equals zero. Consequently, the conventional collision cross section should vanish, cf.~relation~(\ref{vanishing}) above\footnote{More information regarding the phenomenon of infinite photon emission can be found e.g. in~\cite[Sec.~3.2]{Str}.}.

Another manifestation of the IR problem in QED is the so-called infinite Coulomb phase shift. The origin of this problem is the long-range nature of interactions between charged particles, which in low-energy approximation can be described by the classical Coulomb law. It is well-known that the trajectories of the particles interacting via the Coulomb potential are not well-approximated by the trajectories of free particles. The velocities of the particles acquire specific values at the future and past infinity. However, the asymptotic values are approached so slowly that the distance between the particle influenced by the Coulomb potential and a freely moving particle always diverges irrespective of the choice of the initial position and velocity of the free particle. As already mentioned in Sec.~\ref{Dollard-section}, in the case of quantum-mechanical particles interacting via the Coulomb potential the IR problem manifests itself by the nonexistence of the standard M{\o}ller operators. The problem is related to the fact that the wave-function satisfying the Schr{\"o}dinger equation with the Coulomb potential approaches at large distances the free-particle wave-function only up to a logarithmically divergent phase factor, which can be computed e.g. using the Dollard formalism\footnote{For more details regarding the logarithmically divergent Coulomb phase see e.g.~\cite[Sec.~6.2]{Str}.}.

Recall that the scattering amplitudes in massive models of QFT are usually computed with the use of the LSZ reduction procedure~\cite{E2}. The amplitudes of scattering of particles of mass $m$ are expressed in terms of the Fourier transform of an appropriate Green function multiplied by the factors $p_j^2-m^2$ and subsequently restricted to the mass hyperboloids $p_j^2=m^2$, where $p_1,p_2,\ldots$ are the four-momenta of the particles. Let us indicate why the LSZ procedure is not applicable to QED. First, we remark that, as shown in \cite{Bu77}, the standard LSZ limit of the electromagnetic field describing photons exists. However, this is not the case for the Dirac field describing electrons and positrons. In fact, because of logarithmic divergences perturbative corrections to the Green functions in the momentum space are typically more singular in the vicinity of the mass shell than the free Feynman propagator and usually cannot be restricted to the mass shells $p_j^2=m^2$ after multiplication by $p_j^2-m^2$. Consider for example the interacting Feynman propagator, that is the two-point Green function, of the electron. In order to compute this propagator in perturbation theory we have to determine perturbative corrections to the so-called self-energy $\Sigma(p)$ of the electron. After identifying the most singular contributions to $\Sigma(p)$ in the vicinity of the mass shell $p^2=m^2$ and performing a formal resummation we obtain the following asymptotic behavior of the interacting Feynman propagator of the electron
\begin{equation}
 G_2(p) = \frac{\i}{\gamma^\mu p_\mu-m-\Sigma(p)+\i 0} \sim \textrm{const}~\frac{\gamma^\mu p_\mu+m}{(p^2-m^2)^{1-\frac{e^2}{4\pi^2}}}
\end{equation}
for momenta $p$ close to the mass hyperboloid $p^2=m^2$. The above formula suggests that the Feynman propagator of the electron is less singular on the mass shell than the free Feynman propagator and, in particular, it would not have a pole there. As a result, it is expected that in full non-perturbative QED (as of yet non-existent) the standard LSZ procedure would produce vanishing scattering matrix elements for processes involving electrons. On the other hand, the application of the standard LSZ formula in perturbation theory leads to the notorious IR divergencies. Indeed, the terms of the expansion of the propagator in powers of the electric charge $e$ involve logarithmic corrections that diverge on the mass shell and cannot be restricted to the mass shell after multiplication by $p^2-m^2$. Let us note that the above divergences are a manifestation in the perturbative setting of the infra-particle problem, which was described in Sec.~\ref{sec:scattering_electrons_photons}.

\subsection{Attempts to solve IR problem in perturbation theory}\label{sec:origins_IR_perturbative}
 
As explained in the Introduction, the standard pragmatic solution of the problems described in the previous section is to introduce some IR cutoff in the theory, e.g. non-zero photon mass, and apply the LSZ procedure to the theory with the cutoff. The crucial observation, made in~\cite{YFS61}, is that the IR cutoff can be removed at the level of certain infrared safe observables such as inclusive cross sections. Since realistic particle detectors always have finite sensitivity, soft photons with sufficiently small energies may always escape undetected. Thus, the approach seems sufficient for typical practical applications. However, it is not satisfactory from conceptual point of view. Even though the above procedure has been successfully applied to compute low order corrections to various inclusive cross sections, there is still no proof that it works to all orders of perturbation theory. As a matter of fact, there is even no precise mathematical hypothesis that could be proved true or false. More importantly, the physical interpretation of the inclusive cross section is obscure since their construction relies on the use of charged states with sharp momenta that exist only in the theory with some IR regulator. Moreover, when summing over low-energy photons that escape detection one has to include all four polarizations in contradiction to the fact that the photons emitted in radiative processes are transversal\footnote{The physical photons always have transversal polarisation. However, in the standard Gupta-Bleuler quantization of QED additional unphysical photon polarisations are used in the intermediate steps of the construction, see Sec.~\ref{sec:GB}.}. Furthermore, for the procedure to work the summation has to be performed only over soft photons in the outgoing state. We refer the reader to~\cite[Sec.~3.11]{scharf2014} and~\cite[Ch.~17]{St} for some rigorous attempts at defining the inclusive cross sections.

Another strategy to solve the IR problem in QED was put forward by Faddeev and Kulish~\cite{FK70}. Their idea was to construct a certain IR-finite $S$-matrix using the modified scattering theory, which was originally formulated by Dollard~\cite{Do64} to solve the problem of the Coulomb scattering in quantum mechanics. The modified $S$-matrix is constructed by comparing the true dynamics of the system with some non-trivial reference dynamics. The reference dynamics is chosen so that it is explicitly solvable and describes a certain long-range interaction between charged particles that is expected to coincide with the part of the true interaction in QED that is of long-range character and persists for asymptotic times. The Faddeev and Kulish strategy was put on firm mathematical grounds and tested in low orders of perturbation theory in~\cite{duch21infrared} by combining the modified scattering theory with the Bogoliubov method of adiabatic switching of interaction. We give a short overview of this approach in the next section. Note that the ideas of Faddeev and Kulish were used in several simplified models of QED that can be defined non-perturbatively. For the application in the non-relativistic QED see Sec.~\ref{Dollard-section}. Let us also mention the work~\cite{morchio2016infrared} by Morchio and Strocchi who used the modified scattering theory to give a fairly complete analysis of the IR problem in a certain simplified model of QED.

\subsection{Modified scattering theory in perturbative QED}

In this section we present the proposal for the construction of the IR-finite modified scattering matrix in perturbative QED given in~\cite{duch21infrared}. The proposal is based on the ideas of Faddeev and Kulish. However, in contrast to their work~\cite{FK70}, it addresses both the UV and IR problems. The advantage of the method proposed in~\cite{duch21infrared} is a clear separation between the UV and IR problem, thanks to the application of the Bogoliubov method~\cite{bogoliubov1959introduction} of adiabatic switching of the interaction. Let us stress that the results of~\cite{duch21infrared} are incomplete as so far the proposal for the construction of the $S$-matrix was only proved to work in low orders of perturbation theory. 

\subsubsection{Dollard method in perturbation theory}

Let us start by presenting a reformulation of the Dollard strategy, discussed in Subsection~\ref{Dollard-section}, that is applicable to perturbation theory. Consider a certain interacting system of particles modeled by a self-adjoint Hamiltonian $H$ acting in some Hilbert space $\mathcal{H}$. Suppose that the Hamiltonian has the form $H=H_0+eV$, where $H_0$ is the free Hamiltonian and $V$ is some interaction potential. In the modified scattering theory one compares the dynamics generated by the Hamiltonian $H$ with a certain dynamics generated by an appropriate (typically time-dependent) asymptotic Hamiltonian of the form $H_\ras(t) = H_0+e V_\ras(t)$. The interaction potential $V_\ras(t)$ is chosen so that the dynamics generated by $H_\ras(t)$ is simple and captures some relevant features of the full dynamics. For $-\infty<t_2\leq t_1<\infty$ the unitary transformations generated by the time-dependent Hamiltonian $H_\ras(t)$ describing the evolution from time $t_1$ to $t_2$ are explicitly given by\footnote{We assumed that $t_2\leq t_1$ because in subsequent analysis we will only need the unitaries that evolve the state backwards in time.}
\begin{equation}
 U_\ras(t_2,t_1) 
:= \mathrm{\bar Texp}\left(\i\int_{t_2}^{t_1} H_\ras(t)\,d t\right),
\end{equation}
where $\mathrm{\bar Texp}$ is the anti-time-ordered exponential. The modified scattering matrix can be defined by the following weak limit
\begin{equation}
 S_{\mathrm{mod}}:=\lim_{\substack{t_1\to-\infty\\t_2\to+\infty}} 
 U_\ras(0,t_2)\e^{-\i (t_2-t_1) H}  U_\ras(t_1,0).
\end{equation}
In particular, if $V_\ras(t)=0$, then $S_{\mathrm{mod}}$ coincides with the standard scattering matrix. It turns out that, at least formally, the above formula can be rewritten as
\begin{equation}\label{eq:S_mod_formal}
 S_{\mathrm{mod}}=\lim_{\substack{t_1\to-\infty\\t_2\to+\infty}} 
\mathrm{\bar Texp}\left(\i e\int_{0}^{t_2} V_\ras^{\mathrm{I}}(t)\,d t\right)
\mathrm{Texp}\left(-\i e \int_{t_1}^{t_2}d t\, V^{\mathrm{I}}(t)\right)
\mathrm{\bar Texp}\left(\i e\int_{t_1}^{0} V_\ras^{\mathrm{I}}(t)\,d t\right),
\end{equation}
where 
$V^{\mathrm{I}}(t) := \e^{\i t H_0} V\e^{-\i t H_0}$ and 
$V^{\mathrm{I}}_\ras(t) := \e^{\i t H_0} V_\ras(t)\e^{-\i t H_0}$ are the interaction terms in the interaction picture. The r.h.s. of the above formula is interpreted as a formal power series in the coupling constant~$e$ and will serve as a guiding principle in the construction of the modified $S$-matrix in perturbative QED presented in Sec.~\ref{sec:mod_S_QED}.

\subsubsection{Gupta-Bleuler approach to perturbative QED}\label{sec:GB}

Let us recall the basics of the standard Gupta-Bleuler formulation of perturbative QED. The Gupta-Bleuler approach is needed because of the presence of the local gauge symmetries. It can be viewed as a quantization of a modified theory that has more degrees of freedom than QED but, in contrast to QED, has a well defined dynamics (i.e. the Cauchy problem has a unique solution). The modified theory is equivalent to QED when restricted to gauge-invariant observables and states satisfying the Gupta-Bleuler subsidiary condition. The theory is defined in the Krein-Hilbert-Fock space $\mathcal{H}= \Gamma_s(\mathfrak{h}_{\textrm{ph}})\otimes \Gamma_a(\mathfrak{h}_{\textrm{el}})$. Note that the one-particle photon space $\mathfrak{h}_{\textrm{ph}}$ contains photons with four polarizations (including two unphysical ones) and is equipped with the Lorentz-covariant (but not positive-definite) Krein scalar product and a certain positive-definite (but not covariant) scalar product that is used to define the topology. The creation and annihilation operators of photons are denoted by $a_\mu(k)$, $a^*_\mu(k)$, where $\mu=0,1,2,3$ and $k$ is a four-vector on the light-cone. The one-particle electron space $\mathfrak{h}_{\textrm{el}}$ is standard and is equipped with a positive-definite and covariant scalar product. The electron/positron polarization vectors are denoted by $u_a(\sigma,p)$, $v_a(\sigma,p)$ and the creation and annihilation operators -- by $b^*(\sigma,p)$, $b(\sigma,p)$, $d^*(\sigma,p)$, $d(\sigma,p)$, where $\sigma=1,2$. In particular, the free Dirac field is given by
\begin{equation}\label{eq:psi_field}
 \psi_a(x):= \sum_{\sigma=1,2}\int d\mu_m(p) \left(b^*(\sigma,p)u_a(\sigma,p)\e^{\i p\cdot x}+d(\sigma,p)v_a(\sigma,p) \e^{-\i p\cdot x}\right),
\end{equation}
where $d\mu_m(p)$ is the Lorentz invariant measure on the electron mass hyperboloid. The Gupta-Bleuler subsidiary condition requires that $\partial^\mu A_\mu^{-}(x)\Psi=0$ for physical states $\Psi\in\mathcal{H}$, where $A_\mu^{-}(x)$ denotes the negative-energy part of $A_\mu(x)$. For the introduction to the Gupta-Bleuler approach to QED see~\cite[Sec.~8.2]{Str} or \cite[Sec.~2.11]{scharf2014}. 

\subsubsection{Bogoliubov method of adiabatic switching of interaction}\label{sec:Bogoliubov}

Now suppose that $H_0$ is the usual free Hamiltonian in the Fock space $\mathcal{H}$ describing photons and electrons and the interaction term is given by~\eqref{eq:QED_interaction_vertex}. Let us first concentrate on the middle factor under the limit on the r.h.s. of the formula~\eqref{eq:S_mod_formal}. As we mentioned in the Introduction, this expression suffers from both the UV and IR problem. An elegant solution to both problems, proposed by Bogoliubov~\cite{bogoliubov1959introduction}, is to study instead the $S$-matrix with the switching function $g_\epsilon$ defined by
\begin{equation}\label{eq:Bogoliubov_S_matrix}
 S(g_\epsilon):=\mathrm{Texp}\left(\i e\, \int_{\mathbb{R}^4} g_\epsilon(x):\!J^{\mu}(x)A_{\mu}(x)\!:d^4 x \right),
 \qquad
 g_\epsilon(x):=g(\epsilon x),
 \quad\epsilon\in(0,1],
\end{equation}
where $g$ is a fixed Schwartz function on the spacetime such that $g(0)=1$. The IR cutoff $g_\epsilon$ is removed at the end of the construction by taking the so called adiabatic limit $\epsilon\to0$. The adiabatic limit formally corresponds to the limit $t_1\to-\infty,t_2\to+\infty$ in the formula~\eqref{eq:S_mod_formal}. However, note that the switching function regularizes the interaction in both space and time. Crucially, the Bogoliubov method allows to avoid all spurious UV divergences. The coefficients of the formal power series on the r.h.s. of Eq.~\eqref{eq:Bogoliubov_S_matrix} are expressed in terms of the time ordered products of interaction vertices tested with a Schwartz function. As a result, they are automatically well-defined once the time-ordered products are constructed as Schwartz distributions. The construction of the time-ordered products requires the solution of the familiar UV problem and is well-understood, cf. e.g.~\cite{scharf2014}. In purely massive theories one can then define the scattering matrix as a formal power series in $e$ by the adiabatic limit $S=\lim_{\epsilon\to0}S(g_\epsilon)$, cf.~\cite{epstein1976adiabatic}. Because of the IR problem, the adiabatic limit does not exist in QED. However, as we will see in the next section one can define the modified scattering matrix by introducing appropriate asymptotic dynamics formally corresponding to nonzero $V_\ras^{\mathrm{I}}(t)$ in Eq.~\eqref{eq:S_mod_formal}.

\subsubsection{Modified $S$-matrix in perturbative QED}\label{sec:mod_S_QED}

Using the formula~\eqref{eq:S_mod_formal} as a guiding principle we postulate that the modified $S$-matrix in QED is given by the following adiabatic limit $S_{\mathrm{mod}}=\lim_{\epsilon\to0}S_{\mathrm{mod}}(g_\epsilon)$ with\footnote{\label{foot:gauge_inv}The full expression for the modified $S$-matrix involves two extra factors which ensure that the modified $S$-matrix is gauge invariant and the interacting electromagnetic field has desired long-range tail compatible with the Coulomb law. See~\cite{duch21infrared} for details.}
\begin{equation}\label{eq:S_mod_g}
 S_{\mathrm{mod}}(g_\epsilon)=
  S^\ras_{\rout}(g_\epsilon) S(g_\epsilon) S^\ras_{\rin}(g_\epsilon),
\end{equation}
where $S(g_\epsilon)$ is the Bogolibov $S$-matrix introduced in the previous section, $S^\ras_{\rout/\rin}(g_\epsilon)$ are the so-called Dollard modifiers that formally correspond to the first and third factor under the limit on the r.h.s. of the formula~\eqref{eq:S_mod_formal}.

Let us first discuss the construction of the Dollard modifiers $S^\ras_{\rout/\rin}(g_\epsilon)$. By comparing the formulas~\eqref{eq:S_mod_formal} and~\eqref{eq:S_mod_g}, it is clear that, at least formally, the  Dollard modifiers should correspond to the anti-time-ordered exponentials of a certain asymptotic interaction vertices (cf. Sec~\ref{Dollard-section}, where the case of non-relativistic QED is discussed). The dynamics described by the asymptotic interaction should exhibit the same IR problems as the full interaction (i.e. the infinite photon emission and the infinite Coulomb phase discussed in Sec.~\ref{sec:origins_IR_perturbative}) but in contrast to the later should be explicitly solvable. In order to define the asymptotic interaction vertices we replace the spinor current $J^\mu(x)=\,:\!\overline{\psi}(x)\gamma^\mu\psi(x)\!:$ in the interaction vertex of QED $J^\mu(x)A_\mu(x)$ with certain asymptotic currents 
\begin{equation}\label{eq:QED_J_out_in_as}
 J^\mu_{\rout/\rin}(x)
 :=
\int_{H_m} \rho(p) j^\mu_{\rout/\rin}(p;x)\,{ d\mu_m( p) },
 \qquad
 j^\mu_{\rout/\rin}(m v;x):=\pm v^\mu \int_0^\infty \eta\left(x\mp\tau v\right)\,d\tau,
\end{equation}
where the operator $\rho(p)=\sum_{\sigma=\pm}(b^*(p,\sigma)b(p,\sigma)-d^*(p,\sigma)d(p,\sigma))$ can be viewed as the charge density in momentum space, the profile of the charge in position space $\eta$ is a Schwartz function on the spacetime that integrates to one and $v$ is a four-velocity. Note that for $\eta$ equal to the Dirac delta the numerical current $j^\mu_{\rout/\rin}$ coincides with the current of a point particle moving with the velocity $v$ restricted to future/past lightcone\footnote{Such currents are not conserved. To remedy this, one needs to introduce in the definition of $S_\rmod(g_\epsilon)$ two extra factors mentioned in Footnote~\ref{foot:gauge_inv}.}. For our purposes it is crucial that $\eta$ is a smooth function as otherwise some spurious UV divergences appear. It turns out that the IR properties of the asymptotic currents do not depend on the choice of the profile $\eta$ and the large-time future/past asymptotics of $J^\mu_{\rout/\rin}$ and $J^\mu$ coincide. 

In view of the discussion in the previous paragraph, we are led to define the Dollard modifiers as the anti-time ordered exponential of the asymptotic interaction vertices $J^{\mu}_{\rout/\rin}(x)A_{\mu}(x)$. Using the fact that the asymptotic currents commute with themselves and the vector potential we arrive at the following simple explicit formula for the Dollard modifiers
\begin{multline}\label{eq:def_modifiers}
 S^\ras_{\rout/\rin}(g_\epsilon)
 :=
 \exp\left(-\i e\int g_\epsilon(x)\, J^\mu_{\rout/\rin}(x)A_\mu(x)\,d^4 x\right)
 \\ 
 \times\exp\left(\i\frac{e^2}{2}\!\int\! g_\epsilon(x)g_\epsilon(y)
 \,\eta_{\mu\nu}D^D(x-y):\!J^\mu_{\rout/\rin}(x) J^\nu_{\rout/\rin}(y)\!:\,d^4 x d^4 y\right),
\end{multline}
where $\eta_{\mu\nu}$ is the Minkowski metric with the signature $(+,-,-,-)$, $D^D:=\frac{1}{2}(D_{\mathrm{ret}}+D_{\mathrm{adv}})$ and $D_{\mathrm{ret/adv}}$ are the massless retarded/advanced Green functions. The first factor on the r.h.s. of the above formula is the Weyl operator responsible for the generation of a cloud of
photons surrounding massive particles. From perturbative point of view it describes the emission or absorption of a photon by an electron or positron whose momentum is unchanged in this process. The second factor is the exponential of minus the Coulomb phase. Note that for any $\epsilon>0$ the Dollard modifiers are well defined. However, in the limit $\epsilon\to 0$ the Weyl operator converges weakly to zero and the Coulomb phase diverges logarithmically. Let us also stress that no UV problem appears in the construction of $S_{\rout/\rin}(g_\epsilon)$ and the UV problem in the construction of $S(g_\epsilon)$ is solved using the standard techniques, cf. Sec.~\ref{sec:Bogoliubov}.

Because the Bogoliubov $S$-matrix $S(g_\epsilon)$ can only be defined as a formal power series in~$e$, all other factors that appear on the r.h.s. of Eq.~\eqref{eq:S_mod_g} are also interpreted as formal power series. It is expected that the adiabatic limits of all coefficients of the series $S_{\mathrm{mod}}=\lim_{\epsilon\to0}S_{\mathrm{mod}}(g_\epsilon)$ exist. In~\cite{duch21infrared} this was rigorously established for the first two non-trivial coefficients. If the adiabatic limit exists, then $S_{\mathrm{mod}}$ is automatically gauge-invariant\footnote{However, see Footnote~\ref{foot:gauge_inv}.} and thus induces an operator on the physical Hilbert space of states satisfying the Gupta-Bleuler subsidiary condition. Moreover, $S_{\mathrm{mod}}$ is invariant under the action of a certain strongly-continuous unitary representation of the group of translations. The generators of these transformations, i.e. the energy-momentum operators, have the following form
\begin{equation}
 P_\rmod^\mu
 =
 \sum_{s=1,2} \int d\mu_0(k)\, k^\mu \tilde a^*(s,k) \tilde a(s,k) 
 +
 P^\mu_{\mathrm{el}},
\end{equation}
where $\tilde a^*(s,k)$ and $\tilde a(s,k)$ are the creators and annihilators of physical photons and $P^\mu_{\mathrm{el}}$ are the standard energy-momentum operators of free electrons and positrons. The photon creators and annihilators are constructed by taking the past LSZ limit of the modified interacting retarded electromagnetic field and have the following form (in sectors with zero total electric charge)
\begin{equation}\label{eq:asymp_a_qed}
 \tilde a(k,s) = \varepsilon^\mu(s,k) a_\mu(k) +e\,\tilde{\eta}(k) \int_{H_m}d\mu_m(p)\,\rho(p)\frac{\varepsilon^\mu(s,k) p_\mu}{p\cdot k},
\end{equation}
where $\varepsilon^\mu(s,k)$, $s=1,2$, are the transversal photon polarization vectors. Note that creators/annihilators of physical photons do not commute with creators/annihilators of electrons/positrons. In particular,  the Fock vectors $b^*(p,\sigma)\Omega$, where $\Omega$ is the vacuum, are not in the kernel of $\tilde a(k,s)$. In fact, all states with at least one electron/positron contain infinite number of photons. The joint spectrum of $P_\rmod^\mu$ coincides with the closed forward lightcone, $P_\rmod^\mu\Omega=0$ and $(P_\rmod^\mu- k^\mu) \tilde a^*(k,s)\Omega=0$. However, there are no massive one-particle states in the joint spectrum of $P_\rmod^\mu$, which is consistent with the infraparticle nature of the electron, cf. Sec.~\ref{sec:scattering_electrons_photons}. 

\section{\colr Conclusions and outlook}

In this article we outlined some work on the infrared problem of the last two decades.
In the setting of algebraic QFT we discussed Compton scattering in hypercone localized representations introduced
by Buchholz and Roberts. We stressed that the  spacelike asymptotic flux of the electric field (\ref{fluxes}) 
does not exist in such representations. Thus, they are of infravacuum type and the conventional (Wigner)
definition of the electron does not lead to contradictions. In the
complementary class of infraparticle representations the physical electron is a complicated 
composite object consisting of a bare electron and a soft photon cloud correlated with its
velocity. In the setting of non-relativistic QED we surveyed various concrete formulas describing
the electron as an  infraparticle. They were obtained by combined insights from the Haag-Ruelle
scattering theory and the Faddeev-Kulish formalism. As their respective approximating sequences
were proven to converge, they capture correctly the asymptotic dynamics.
The Faddeev-Kulish   approach has also been rigorously implemented in perturbative QED. 
Here it gave a candidate scattering matrix (\ref{eq:S_mod_g}) which is finite both in the infrared and ultraviolet regime {\colg at least} in low
orders of perturbation theory.

Verifying the finiteness of the latter scattering matrix to all orders of perturbation theory and checking its
consistency with the YFS algorithm for inclusive cross sections is an important direction for future research. Arguably, such a
result could be called a solution of the infrared problem in QED.   As a preparatory step, such a  comparison can be attempted
in non-relativistic QED:  Formulas~(\ref{CFP-representation}), (\ref{BDG}) for the physical electron can be used to define
Compton scattering states and study the corresponding scattering matrix elements. As mentioned above, generalization of
these formulas to several electrons, aiming at Coulomb scattering, appears to be within reach of existing methods. Construction
of scattering states of several electrons is also an interesting open problem in the infravacuum setting of Buchholz and Roberts, 
cf. Sec.~\ref{sec:scattering_electrons_photons}. This requires further developments of superselection theory, including a construction of charge carrying fields of the electron. Such fields are relativistic counterparts of the renormalized creation operators~(\ref{electron-creation-op-ren}), which are crucial ingredients of multi-electron states. 

Summing up, the infrared problem still  {\colg provides} a number of interesting and feasible
research questions, which touch upon the nature of light and electrically charged particles.
 These problems  involve diverse mathematical tools, such as  $C^*$-algebras 
and their representations, spectral theory for embedded eigenvalues and  various renormalization 
techniques.  Therefore, we expect that the infrared problem  will remain an active field of mathematical physics.



\begin{thebibliography}{LNT2}
\bibitem[E]{E} The references marked with an E in front of their numbers are articles solicited for this 
encyclopedia. The titles may change.
\end{thebibliography}
\end{document}